\documentclass[prl,reprint,aps,superscriptaddress]{revtex4-1}

\usepackage{amsmath}	
\usepackage{amssymb}	
\usepackage{graphicx}
\usepackage{dcolumn}
\usepackage{bm}
\usepackage{hyperref}
\usepackage{url}
\usepackage{subcaption}

\usepackage{appendix}

\newcommand\mnras{MNRAS}
\newcommand\ssr{Space Sci. Rev.}     
\newcommand\solphys{Sol.~Phys.}      
\newcommand\aap{A\&A}                
\newcommand\lb{\langle}              
\newcommand\rb{\rangle}               

\begin{document}

\title{Sustained turbulence and magnetic energy in non-rotating  shear flows}

\author{Farrukh Nauman}
\email{nauman@nbi.ku.dk}
\affiliation{Department of Physics and Astronomy, University of Rochester, Rochester, NY 14627, USA.}
\affiliation{Niels Bohr International Academy, The Niels Bohr Institute, Blegdamsvej 17, DK-2100, Copenhagen \O, Denmark.}
\author{Eric G. Blackman}
\email{blackman@pas.rochester.edu}
\affiliation{Department of Physics and Astronomy, University of Rochester, Rochester, NY 14627, USA.}
	
\date{\today}
	
\begin{abstract}
From numerical simulations, we show that non-rotating magnetohydrodynamic shear flows are unstable to finite amplitude velocity perturbations and become turbulent, leading to the growth and sustenance of magnetic energy, including large scale fields. This  supports the concept that sustained magnetic energy from turbulence is independent of the driving mechanism for large enough magnetic Reynolds numbers.
\end{abstract}

\maketitle

\section{Introduction}
Shear flows  are common in nature, both rotating and non-rotating. Rotation is essential when angular momentum support causes the shear, and most studies of field growth have focused on the former. But shear flows in which rotation is inessential are also ubiquitous.
Examples from astrophysics occur near the interface of outflows propagating into ambient media \cite{Liang2013}, near the interface of outward and inward convective plumes in disks or stars \cite{Miesch2009} and between turbulent eddies in galaxies or cluster gas \cite{Bruggen2015}. For some azimuthal shear flows, the role of rotation may also be minimal \cite{2000SoPh..194..189C}. Magnetic fields are common in all of these contexts. Whether linear shear flows can generate turbulence, amplify magnetic energy, or even produce large scale fields are all questions that interface into the long standing questions of magnetic field amplification in astrophysics and identifying the minimum conditions needed for field amplification in magnetohydrodynamics (MHD) (for a review of dynamo theory, see \cite{2012SSRv..169..123B}).
    
While amplification of magnetic fields in stochastically forced flows with and without shear has been demonstrated, questions of  nonlinear stability and magnetic field sustenance in non-rotating magnetized shear flow in three dimensions have received  little attention (for two dimensions, see \cite{2014PhRvE..89d3101M}). Purely hydrodynamic linearly stable shear flows do indeed transition into a turbulent state for a variety of hydrodynamic systems at high Reynolds number, $Re$ (for example, plane Couette flow (PCF): \cite{manneville2015}, pipe flow: \cite{mullin}) and without a net magnetic flux, the perturbed velocities are affected by magnetic field fluctuations only to second order. The linear stability problem thus reduces to that of hydrodynamic PCF but can the resulting turbulence sustain magnetic energy? 

Previously, Ref. \cite{1996ApJ...464..690H} found  that while  magnetized linear shear flows do indeed exhibit flow turbulence, magnetic energy  was found to decay. This  was interpreted to suggest that linear shear flows may be intrinsically unable to grow fields in the absence of rotation. While the Coriolis force stabilizes hydrodynamic Keplerian shear flow (Rayleigh criterion, e,g. \cite{2015MNRAS.448.3697S}), Ref. \cite{1996ApJ...464..690H} emphasized that magnetized Keplerian shear is linearly unstable to the magnetorotational instability (MRI) (\cite{1959velikhov, 1960PNAS...46..253C, 1991ApJ...376..214B, *BNreview,2015MNRAS.448.3697S}) which does sustain growth. However, Ref. \cite{1996ApJ...464..690H} employed low resolution ideal MHD simulations, and this problem of linear shear was not studied for convergence. Their conclusions also lead to a cognitive dissonance: if turbulence from linear shear flows were distinctly unable to sustain magnetic energy, it would contradict a lesson from stochastically forced turbulence where saturated magnetic energy achieves near equipartition with turbulent kinetic energy for large enough magnetic Reynolds number, $Rm$ (\cite{sch2004, haugen2004}). We are thus motivated to revisit this non-rotating magnetized shear problem with  more comprehensive simulations.
 
There is also long standing interest in understanding the role of shear in the generation of large scale magnetic fields (e.g., \cite{2008PhRvL.100r4501Y, 2009PhRvE..79d5305S, 2011ApJ...738...66S, 2013Natur.497..463T, sridhar2014, squire2015PRL, 2015ApJ...813...52S}). 
Ref. \cite{2008PhRvL.100r4501Y}  was the first to show numerically using a shearing box, that the combination of non-helical stochastic forcing plus linear shear can lead to large scale dynamo in a shearing box. The forcing in the simulations of Ref. \cite{2008PhRvL.100r4501Y} 
is such that the stochastic power input was much stronger than the shear forcing, and scale separation was achieved through the use of large vertical domains. But the aforementioned studies of non-rotating linear shear and large scale magnetic field growth have employed the additional stochastic forcing as the primary source of turbulence. This contrasts our present work.
 
In this paper, we study the nonlinear stability of non-rotating magnetized shear flow using a suite of numerical simulations in a shearing box, without any additional stochastic forcing. We explore the sustenance of turbulent state as well as the creation of large scale magnetic fields. The behavior of velocity and magnetic fields is studied as a function of box size and the dissipation coefficients. Most significantly we find that linear shear flows unstable to turbulence do indeed sustain magnetic energy for large enough magnetic Reynolds numbers.  
	
\section{Methods and Results} We perform direct numerical simulations (DNS) employing a shearing box setup (with no other forcing) to study non-rotating magnetized linear shear flow using the pseudospectral code \textsc{snoopy} \footnote{\url{http://ipag.osug.fr/~lesurg/snoopy.html}} (\cite{2011A&A...528A..17L}). We define $Re = L_x^2 S/\nu$, $Rm = L_x^2 S/\eta$ 
where $L_x = 1$ is the size of domain in the `x' direction, $\eta$ is the magnetic diffusivity, $\nu$ is the kinematic viscosity and the shear parameter, $S = 1$. We set the magnetic Prandtl number $Pm = Rm/Re = 1$ for all runs. We initialize our simulations with zero net magnetic flux ${\bm B}_{\text{ini}} = B_0 \sin k_x x {\bm e}_z$ (where $B_0 = 0.035$) and apply finite amplitude perturbations ($LS$) to large scales in the velocity \footnote{We do not conduct a detailed study on the critical amplitude required to trigger and sustain turbulence and simply use velocity perturbations of amplitude $LS$ for all of our simulations.}. The shear profile ${\bm V}_{\text{sh}} = - Sx {\bm e}_y$ is subtracted out of the total velocity and the velocity the code solves for is $\bm{V} = {\bm V}_{\text{total}} - {\bm V}_{\text{sh}}$:
\begin{gather}
\frac{\partial \bm{V}}{\partial t} + V_{sh} \frac{\partial \bm{V}}{\partial y} + \nabla \cdot (\bm{V} \bm{V} + \bm{T}) = -S V_x \bm{e}_y + \nu \nabla^2 \bm{V}, \label{eq:NS} \\
\frac{\partial \bm{B}}{\partial t} = \nabla \times (\bm{V} \times \bm{B}) + \eta \nabla^2 \bm{B}, \label{eq:induc}\\
\nabla \cdot \bm{V} = 0, \quad \nabla \cdot \bm{B} = 0, \notag
\end{gather}
where $\bm{T} = (p+B^2/2)\bm{I} - \bm{B}\bm{B}$. The shear time unit is $1/S$. 

\begin{figure}
	\centering
	\includegraphics[width=0.45\textwidth]{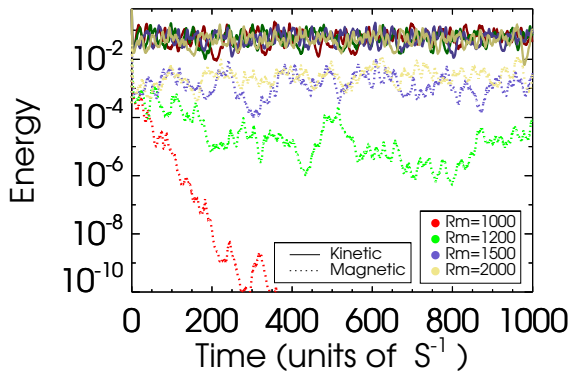}
	\caption{Time history plot for different runs. The kinetic energy is represented by solid lines that are all at the top of the plot. The dotted lines represent magnetic energy. The bottom most line (red, dotted) represents $Re=Rm=1,000$ that decays. The $Re=Rm=1,200$ (green, dotted) line is second from bottom, $1,500$ (blue, dotted) third from bottom and $2,000$ (khaki, dotted) for a domain size of $1\times 2\times 1$ with $32\times 64\times 32$ zones. Both kinetic and magnetic energy are normalized by $S L_x^2$.}
	\label{fig:trans}
\end{figure}

\subsection{Critical $Re$ and $Rm$} In figure \ref{fig:trans}, we plot the time history of the kinetic and magnetic energies for runs with resolution $32\times 64\times 32$. We identify three distinct regimes: (1) $Re<Re_{\text{crit}}$: the flow remains largely laminar and the initial perturbations die off, (2) $Re>Re_{\text{crit}}$ but $Rm<Rm_{\text{crit}}$: kinetic energy grows and sustains for sometime while the magnetic energy decays immediately after reaching the saturation state, (3) $Re>Re_{\text{crit}}$ and $Rm>Rm_{\text{crit}}$: Both kinetic and magnetic energy sustain growth. Fig. \ref{fig:trans} shows that $Rm_{\text{crit}} \sim 1,200$. We estimated $Re_{\text{crit}}\sim 750$ for both hydrodynamic and MHD runs, which is consistent with the value found in the hydrodynamic simulations of PCF \cite{manneville2015} (note that the definition typically used in PCF literature is a factor of 2 smaller than our definition). These critical values have also been verified at a higher resolution of $64\times 128\times 64$. Note that the finite lifetime of turbulence as seen in the kinetic energy of $Re=Rm=1,000$ (red) run in fig. \ref{fig:trans} is consistent with what has recently been found in hydrodynamic shear flow experiments \cite{hof2006Nat}. Ref. \cite{manneville2015} suggests that turbulence in linear shear flows in small domains ($L_y,L_z \sim L_x$) exhibit transient chaos, while large aspect ratio domains $L_y,L_z \gg L_x$ would instead abruptly transition into steady turbulence. We do not explore  extended domains herein so  our results would represent a lower limit on the robustness of turbulence.
\begin{figure}
	\centering
	\includegraphics[width=0.45\textwidth]{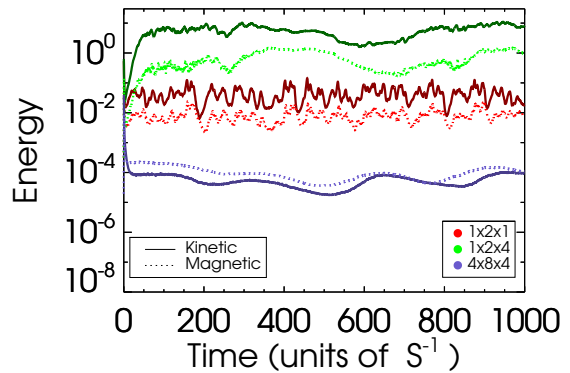}
	\caption{Same as \ref{fig:trans} for $1\times 2\times 1$ (red, middle two lines), $1\times 2\times 4$ (green, top two lines), $4\times 8\times 4$ (blue, bottom two lines) at $Rm = 10,000$.}
	\label{fig:histcom}
\end{figure}

\subsection{Domain Size effects} We plot the ratio of kinetic energy to magnetic energy in fig. \ref{fig:histcom} for three different domain sizes (see table \ref{tab:runs} for description) $1\times 2\times 1$ ($64^3$), $1\times 2\times 4$ ($64\times 64\times 256$) and $4\times 8\times 4$ ($256^3$). The table also has data for a higher resolution run for the domain $1\times 2\times 1$, which suggests convergence for this aspect ratio. The magnetic energy is nearly an order of magnitude smaller than kinetic energy in $1\times 2\times 1$ and $1\times 2\times 4$ whereas in the the largest domain, $4\times 8\times 4$ the two are nearly equal. For the range of domains studied so far, this energy ratio therefore depends not only on the aspect ratio $L_z/L_x$ but increases  with box size for a fixed aspect ratio. 

\begin{table}
	\begin{tabular}{ | c | c | c | c | c | }
		\hline
		Domain Size & Resolution & $\frac{\overline{[B^2]}}{\overline{[V^2]}}$ & $\frac{\overline{\partial{\lb\mathcal{E}_x\rb} /\partial z}^{z+} } {\overline{\lb SB_x\rb}^{z+}}$ & Rm \\ \hline
		$1\times 2\times 1$ & $64\times 64\times 64$ & $0.204$ & $0.017$ & $10,000$ \\ \hline
		$1\times 2\times 1$ & $128\times 128\times 128$ & $0.211$ & $0.007$ & $10,000$ \\ \hline
		$1\times 2\times 4$ & $64\times 64\times 256$ & $0.062$ & $0.002$ & $10,000$ \\ \hline
		$4\times 8\times 4$ & $256\times 256\times 256$ & $2.413$ & $0.011$ & $40,000$ \\ \hline
		\hline
	\end{tabular}
	\caption{Description of the three runs analyzed herein, with one higher resolution run done for convergence test. The third column is the ratio of magnetic to kinetic energies $\overline{[B^2]}/\overline{[V^2]}$ while the fourth column shows that the shear dominates the EMF derivative term $\partial{\overline{\lb\mathcal{E}_x\rb}}/\partial z$ in the induction equation for the y-component of the magnetic field ($[Q] = $ volume average of $Q$, $\overline{Q} = $ time average of $Q$ from $100 S^{-1}$ to $200 S^{-1}$, $Q^{z+}$ represents the vertical average from $z=0$ to $z=+L_z/2$, and $xy$ average of Q is represented by $\lb Q\rb$.). The last column lists the magnetic Reynolds number for the three different runs.}
	\label{tab:runs}
\end{table}

\begin{figure*}
	\centering
	\includegraphics[width=0.925\textwidth]{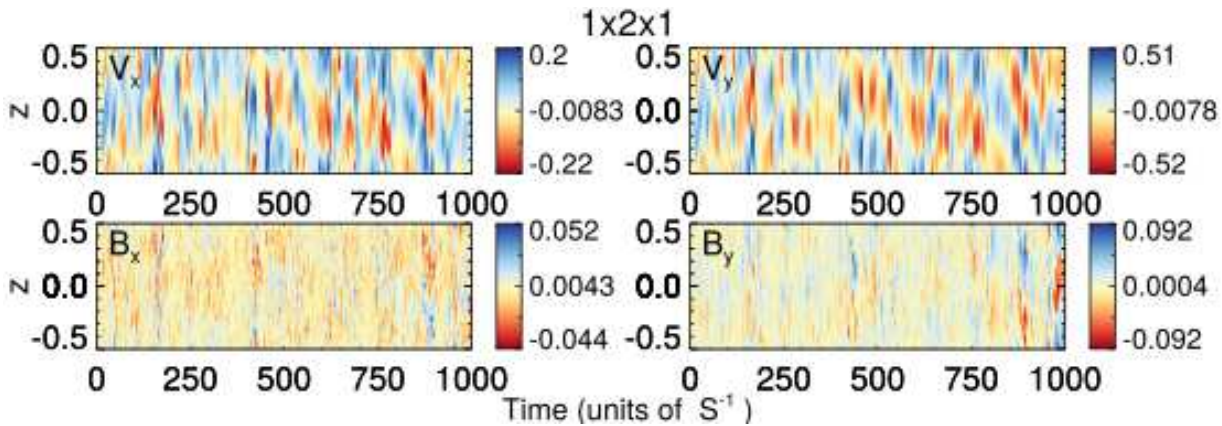}
	\caption{$xy$ averaged $\lb V_x\rb$, $\lb V_y\rb$, $\lb B_x\rb$ and $\lb B_y\rb$ for the run $1\times 2\times 1$ for the first $1000$ shear times. The x-axis is the time ($1/S$ units) and the y-axis is the vertical domain size in $L$ units.}
	\label{fig:1x2x1}
\end{figure*}

The velocity profiles $\lb V_x\rb$, $\lb V_y\rb$ and the magnetic field profiles $\lb B_x\rb$, $\lb B_y\rb$ averaged over $xy$ are plotted for the run with $1\times 2\times 1$ domain in fig. \ref{fig:1x2x1}. Unlike the magnetic fields, we see that while the velocities are dominated by a sinusoidal profile in `z' that varies in time. Furthermore, the $\lb B_x\rb$ profile is more noisy than $\lb B_y\rb$. The simulation began with a shear profile ${\bm V}_{\text{sh}} = - Sx {\bm e}_y$, and eventually reached a steady state with additional shear in the $z-$direction for the $xy$ averaged velocity fields, $\lb \bm{V}\rb_{\text{ver. sh}} \sim \sin k_z z ({\bm e}_x + {\bm e}_y)$. This structure is a generic feature of hydrodynamic shear flows at and just above $Re_{\text{crit}}$. Recent work on the transition to turbulence suggests that as the domain size and $Re$ are increased, these structures disappear into `featureless' turbulence \cite{manneville2015,barkley2005PRL}. 
\begin{figure*}
	\centering
	\includegraphics[width=0.9\textwidth]{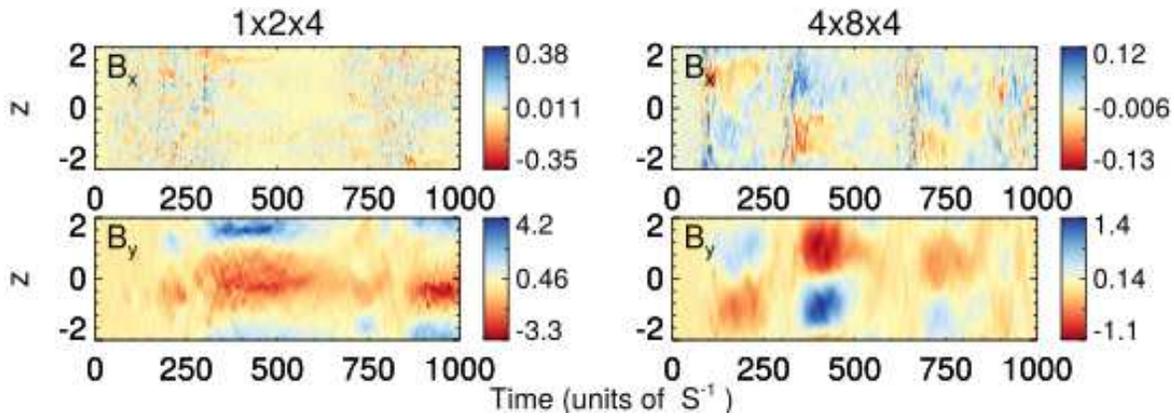}
	\caption{$xy$ averaged $\lb B_x\rb$ and $\lb B_y\rb$ for the runs $1\times 2\times 4$ (left column) and $4\times 8\times 4$ (right column) for the first $1000$ shear times. The x-axis is in the units of $1/S$, while the y-axis represents the vertical domain in $L$ units. Note a strong anti-correlation between the $\lb B_x\rb$ and $\lb B_y\rb$ for both runs. }
	\label{fig:Lz4}
\end{figure*}

To explore possible generation of an organized magnetic field, we plot the magnetic field profiles for the two larger domains with $L_z=4$: $1\times 2\times 4$ and $4\times 8\times 4$ in fig. \ref{fig:Lz4}. The `y' component of the magnetic field, $\lb B_y\rb$ seems to be very coherent in both runs and $\lb B_x\rb$ in the $4\times 8\times 4$ run displays considerable large scale organization compared 
to the corresponding plot for $1\times 2\times 4$. The magnetic fields are strongly correlated with the velocity fields (see Appendix)
for $4\times 8\times 4$ since all of them have sinusoidal structure on the box scale. More interestingly, the magnetic to kinetic energy ratio is nearly unity (see table \ref{tab:runs}). This is in contrast to the $1\times 2\times 4$ run where the magnetic energy is more than an order of magnitude smaller than the kinetic energy and the $\lb B_x\rb$ profile seems to be very noisy. The cross helicity of the $1\times 2\times 4$ run is close to unity for a significant duration, while that of $4\times 8\times 4$ is smaller in comparison and fluctuates about zero suggesting that the $1\times 2\times 4$ run is dominated by $k_z=1$ mode (see Appendix for plots).

The spatiotemporal profile of $\lb B_y \rb$ in fig. \ref{fig:Lz4} suggests the existence of a cycle period and thus a large scale dynamo. In the $xy$ averaged induction equation:
\begin{align}
\frac{\partial \lb B_x\rb}{\partial t} &= - \frac{\partial}{\partial z} \lb V_z' B_x' - V_x' B_z' \rb + \eta \frac{\partial^2}{\partial z^2} \lb B_x\rb \notag \\
\frac{\partial \lb B_y\rb}{\partial t} &= S \lb B_x\rb + \frac{\partial}{\partial z} \lb V_y' B_z' - V_z' B_y' \rb + \eta \frac{\partial^2}{\partial z^2} \lb B_y\rb
\label{eq:ind}
\end{align}
the only terms that can contribute to the right side of the $\lb B_y\rb$ equation are the EMF term $\partial_z \lb\mathcal{E}_x\rb$ (where $\lb\bm{\mathcal{E}}\rb = \lb\bm{V'}\times\bm{B'}\rb$, $V'$ and $B'$ are fluctuations resulting from $xy$ averaging) and the `Omega' term $S\lb B_x\rb$ as seen in eq. \ref{eq:ind} (the mean field term contribution, $\partial (\lb V\rb \times \lb B\rb)_i/\partial z \sim 0$, where $i=x,y$). We estimated the former to be roughly two orders of magnitude smaller than the shear term (table \ref{tab:runs}) 
\begin{figure}[htb]
	\centering
	\includegraphics[width=0.45\textwidth]{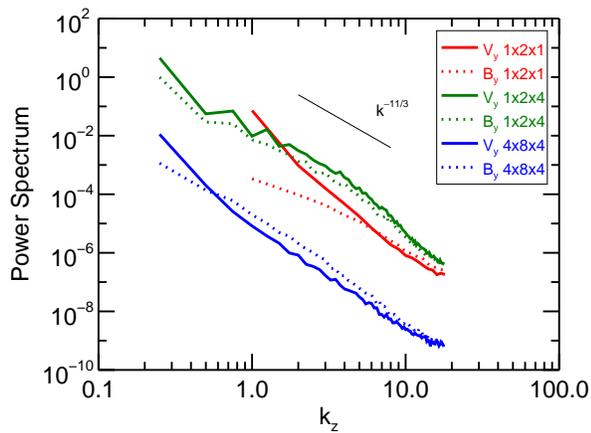}
	\caption{Power spectra of $xy$ averaged azimuthal velocity $\lb V_y\rb$ and magnetic field $\lb B_y\rb$ averaged from $501$ to $600$ shear times as a function of $k_z$ (in units of $(2\pi n_z/L_z)$). The top (green) two lines represent $1\times 2\times 4$, while the middle (red) two represent $1\times 2\times 1$ and $4\times 8\times 4$ (blue) lines are at the bottom. The power spectra are normalized such that $\sum_{k_z} |\lb Q\rb(k_z)|^2 = [Q^2]$. }
	\label{fig:spec}
\end{figure}

We plot the power spectra of the velocity and magnetic field averaged over $xy$ and Fourier transformed in `z' in fig. \ref{fig:spec}. 
The predominance of the $k_z=1$ (in units of $(2\pi n_z/L_z)$) mode is consistent with the nearly sinusoidal profile of $\lb V_y\rb$ seen in fig. \ref{fig:1x2x1} and $\lb B_y\rb$ in  \ref{fig:Lz4}. The velocity spectrum for $1\times 2\times 4$ seems to follow the 1D Kolmogorov scaling $k^{-11/3}$ for intermediate wavemodes, which would imply that these modes represent the inertial range. However, the velocity power spectrum peaks at the box scale, which could be due to strong 2D vortex structures \cite{2011NatPh...7..321X}. The velocity structures in the smallest domain $1\times 2\times 1$ do not seem to follow the 1D Kolmogorov scaling, while the largest domain $4\times 8\times 4$ has a steeper power law behavior for the most part but appears to have a flatter power law spectrum closer to the dissipation scale.

Since the velocity fields are dominated by box scale structures, it becomes a subtle matter to define and distinguish small vs. large scale dynamos \cite{Ebrahimi2016} or system scale dynamo \cite{tobias2011}. An additional caveat is that we are using periodic boundaries and thus the large power observed in box scale structures is an indication that the boundary conditions are strongly influencing the flow dynamics. It remains to be explored whether magnetic fields in such high $Re$ turbulent flows with a featureless velocity profile would show large scale organization. We do note that recent analytic theory for large scale field growth in shearing boxes shows that rotation is not necessary for dynamo action when a source of velocity fluctuations is present in a shear flow \cite{Ebrahimi2016}. Our simulations satisfy their minimum sufficient conditions, although we focus on $xy$ averaging rather than the $yz$ averaging of their case.

\section{Conclusions} 
Using high resolution 3-D simulations of a shearing box with a pseudospectral code, we have demonstrated numerically for the first time that not only does shear driven turbulence sustain for high enough $Re$, but this turbulence amplifies and sustains magnetic energy when $Rm$ is large enough. This contrasts the work of Ref. \cite{1996ApJ...464..690H}  who did not identify sustained growth in magnetic energy because their $Rm$ was below the critical value we have found. The turbulence emergent in our simulations is self-sustained by the linear shear and thus distinct from a different class of work that employed stochastic forcing in addition to the non-rotating shear \cite{2008PhRvL.100r4501Y, 2009PhRvE..79d5305S, 2011ApJ...738...66S, 2013Natur.497..463T, sridhar2014, squire2015PRL, 2015ApJ...813...52S}. Structures in both velocity and magnetic fields at the largest scales are seen in our largest domains and we have identified the EMF terms that sustain the latter. Whether the velocity structures break into featureless turbulence at even higher Reynolds numbers and domain sizes remains to be explored, but the minimum ingredients  derived for large scale field growth \cite{Ebrahimi2016} are met.  


\paragraph*{Acknolwedgements:-}
We thank F. Ebrahimi,  P. Bhat, and S. Tobias for related discussions. The simulations reported on in this paper were done on the Blue Streak cluster hosted by the Center for Integrated Research Computing at the University of Rochester. FN acknowledges funding from the European Research Council under the European Union’s Seventh Framework Programme (FP/2007-2013) under ERC grant agreement 306614. EB acknowledges support from  grants HST-AR-13916.002 and NSF-AST1515648. 


\section{Appendix}
\begin{figure}
	\centering
	\includegraphics[width=0.45\textwidth]{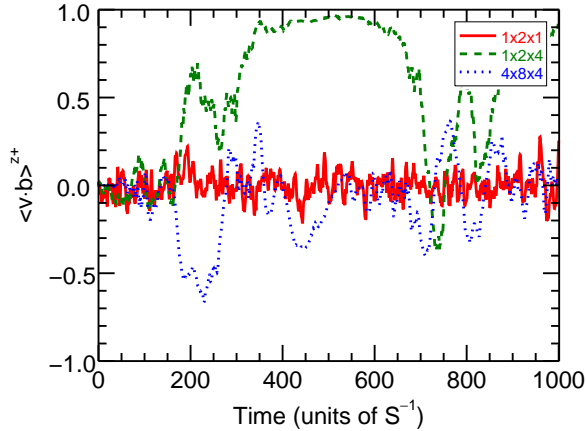}
	\caption{Time evolution of horizontally averaged cross helicity $\lb {\bf v}\cdot {\bf b}\rb^{z+}$ vertically averaged from $z=0$ to $z=+L_z/2$ plotted for $1\times 2\times 1$ (red, solid), $1\times 2\times 4$ (green, dashed), $4\times 8\times 4$ (blue, dotted) for the first $1000 S^{-1}$. }
	\label{fig:north}
\end{figure}
\begin{figure}[hb]
	\centering
	\includegraphics[width=0.45\textwidth]{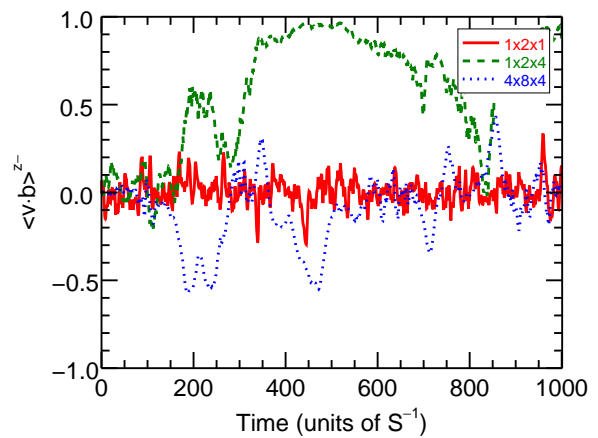}
	\caption{Same as fig. \ref{fig:north} but for $\lb {\bf v}\cdot {\bf b}\rb^{z-}$ vertically averaged from $z=-L_z/2$ to $z=0$. }
	\label{fig:south}
\end{figure}
\begin{figure}[hb]
	\centering
	\includegraphics[width=0.45\textwidth]{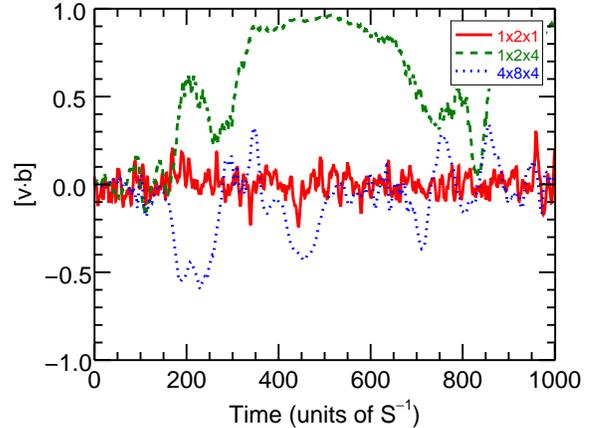}
	\caption{Same as fig. \ref{fig:north} but for  $[{\bf v}\cdot {\bf b}]$ volume averaged. }
	\label{fig:total}
\end{figure}

\begin{figure*}
	\centering
	\includegraphics[width=0.9\textwidth]{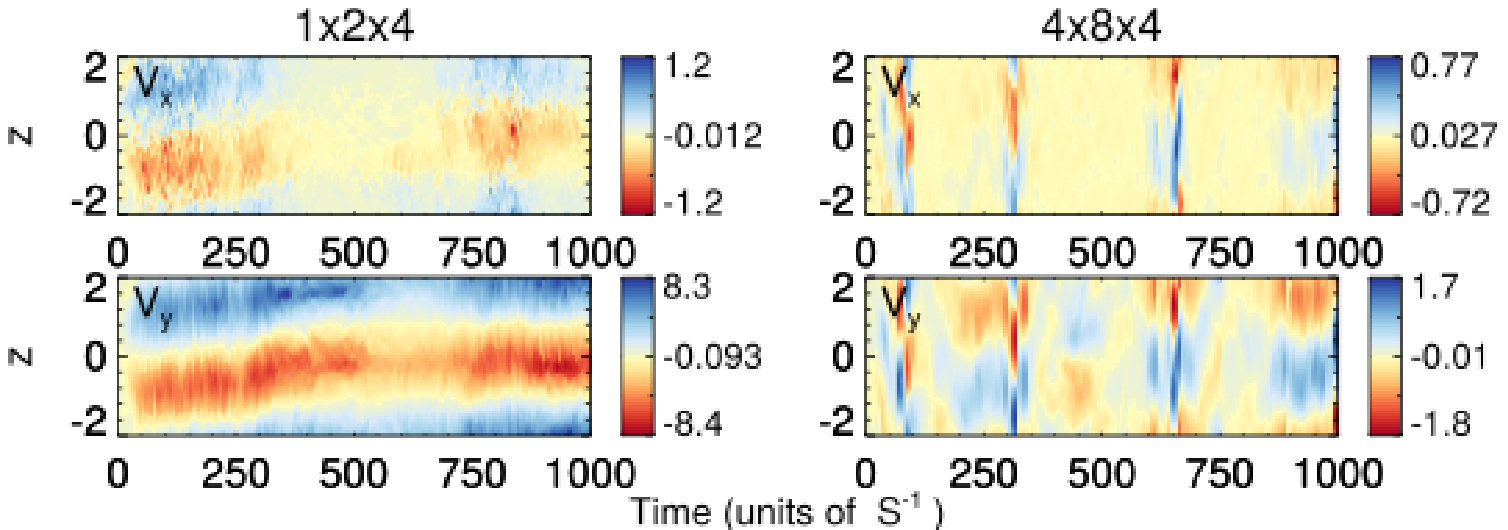}
	\caption{$xy$ averaged $\lb V_x\rb$ and $\lb V_y\rb$ for the runs $1\times 2\times 4$ (left column) and $4\times 8\times 4$ (right column) for the first $1000$ shear times. Compare fig. 4 from the main paper.}
	\label{fig:vel}
\end{figure*}

We plot the cross helicity averaged from $z=0$ to $z=+L_z/2$ represented by $\lb {\bf v}\cdot {\bf b}\rb^{z+}$ ($\lb {\bf v}\cdot {\bf b}\rb^{z-}$: $z=-L_z/2$ to $z=0$; $[{\bf v}\cdot {\bf b}]$: volume average) for the three runs in figs. \ref{fig:north}, \ref{fig:south}, \ref{fig:total}. It appears that the $1\times 2\times 4$ run has a cross helicity dominated by the largest mode and thus has nearly maximal cross helicity with the same sign for different vertical sections of the box for a considerable duration of time. This is further supported by the velocity profiles of this run that are also attached in fig. \ref{fig:vel}. The velocity profiles for this run are all seemingly locked into $k_z=1$ state, similar to the magnetic field profiles in fig. 4 of the main text. Of the 3 runs, the $1\times 2\times 4$ is the only one with a dominant vertical extent ($L_z/L_x > 1$). The large vertical extent seems to be required for the appearance of this large scale dominant mode.


\bibliography{general}
\bibliographystyle{apsrev4-1}

\end{document}